\documentclass[conference]{IEEEtran}
\IEEEoverridecommandlockouts
\usepackage{cite}
\usepackage{amsmath,amssymb,amsfonts}
\usepackage{algorithmic}
\usepackage{graphicx}
\usepackage{textcomp}
\usepackage{xcolor}
\usepackage{url}
\usepackage{subcaption}
\usepackage{multirow}
\usepackage{array}
\usepackage{booktabs}
\usepackage{float}
\def\BibTeX{{\rm B\kern-.05em{\sc i\kern-.025em b}\kern-.08em
    T\kern-.1667em\lower.7ex\hbox{E}\kern-.125emX}}
\begin{document}

\title{Person Detection in Collaborative Group Learning Environments
       Using Multiple Representations}
\author{Wenjing Shi$^1$,
	Marios S. Pattichis$^1$, Sylvia Celed\'on-Pattichis$^2$
	and Carlos L\'opezLeiva$^2$ \\
	$^1$ 
	\textit{image and video Processing and Communications Lab
	(\url{ivpcl.unm.edu})} \\
	\textit{Dept. of Electrical and Computer Engineering}\\
	University of New Mexico, United States.\\		
	$^2$
	\textit{Dept. of Language, Literacy, and Sociocultural Studies}\\
	University of New Mexico, United States.\\
	\{wshi, pattichi, sceledon, callopez\}@unm.\fontfamily{ptm}edu
}

\maketitle

\begin{abstract}
We introduce the problem of detecting a group of
   students from classroom videos.
The problem requires the detection of students
     from different angles and the separation
     of the group from other groups in long videos
     (one to one and a half hours).
     
We use multiple image representations
     to solve the problem.     
We use FM components to separate each group from background groups,
     AM-FM components for detecting the back-of-the-head, and
     YOLO for face detection.
We use classroom videos 
     from four different groups to validate
     our approach.
Our use of multiple representations
     is shown to be significantly more accurate than
     the use of YOLO alone.     
\end{abstract}


\section{Introduction}
We study the problem of person detection in 
   collaborative learning environments' videos.
The problem has some unique challenges associated
    with detecting students sitting around a table.
 
\begin{figure}[!b]
	\centering
	\includegraphics[width=0.48\textwidth]{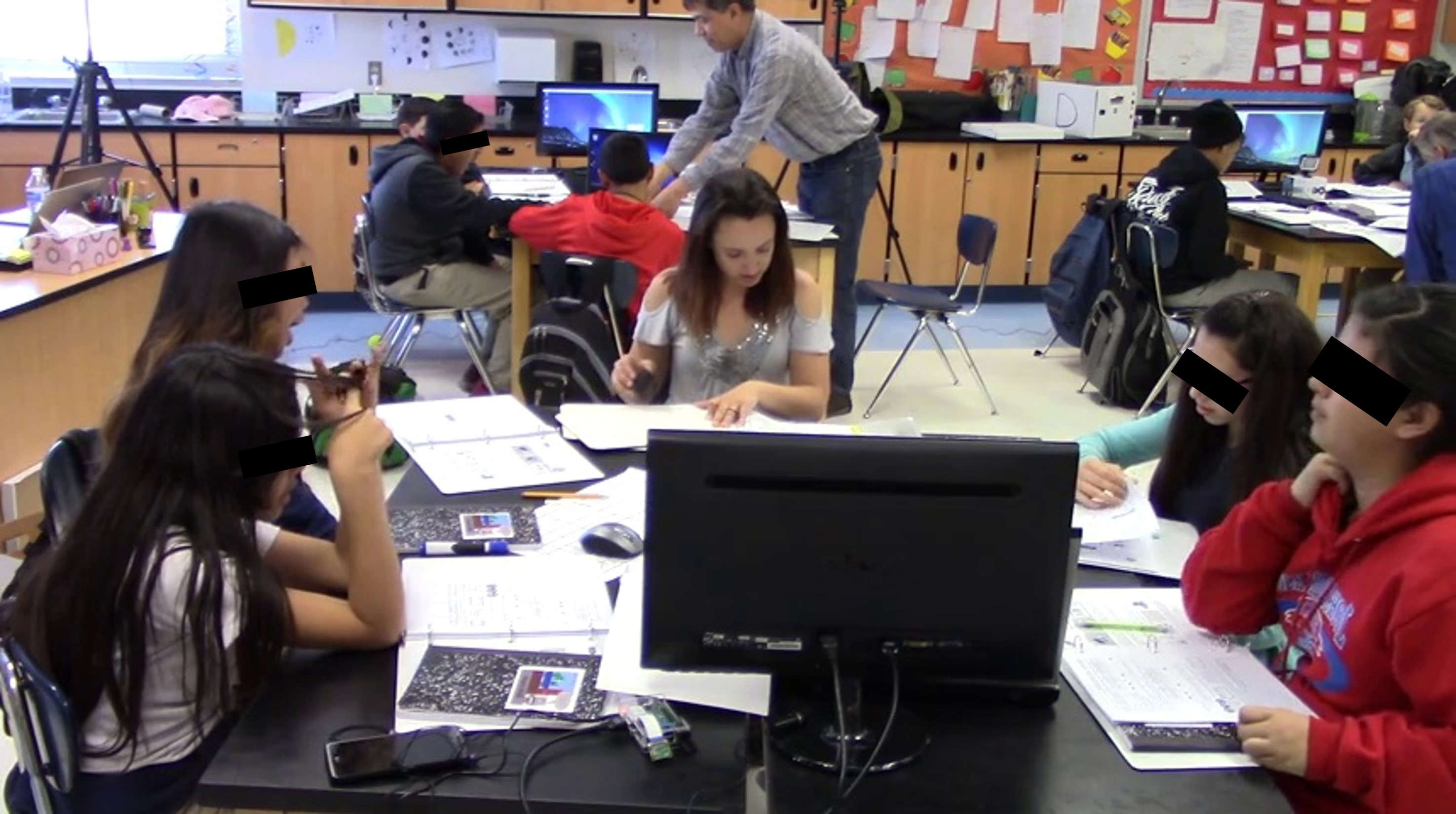}
	\caption{An example of a group of students participating in 
		     a collaborative learning environment.} 
	\label{fig:AOLMESample}
\end{figure}

We present an example of the collaborative learning environment in Figure \ref{fig:AOLMESample}. 
The students' faces are imaged from different angles. They are at different distances from the camera. In many cases, the faces are not visible. Furthermore, there are other groups in the background. 
For the purposes of this paper, 
    we are only interested in detecting students 
    that are sitting around the table
    that is closest to the camera.
All other students are not to be included in the analysis.

Currently, human detection methods are dominated by neural network methods.
As an example, in \cite{nikouei2018real}, 
    the authors used a lightweight Convolutional Neural Network (L-CNN) to detect 
    humans in surveillance video frames. 
In another example, in \cite{zhang2017joint}, the authors 
    used a multi-stream multitask deep network for joint human detection and head poses estimation in RGB-D videos.

We also provide a summary of prior research on classroom videos.
In \cite{shi2016human}, we considered using K-NN classifiers with AM-FM representations
    for person detection.
In \cite{shi2018robust}, the combination of color and FM representations was considered
    for face detection.
In \cite{shi2018robust}, back-of-the-head detection was performed using AM-FM representations.
In \cite{shi2018dynamic}, the method in \cite{shi2018robust} was extended to detect where
   the students were looking.
The importance of FM representations for face detection was further documented
   in \cite{tapia2020importance}.
In \cite{shi2021talking}, the authors used head detection to detect talking activities.
In \cite{teeparthi2021fast}, the authors developed methods for hand detection.
In \cite{sanchez2021bilingual}, the authors considered the use of YOLO \cite{redmon2018yolov3} for head
    detection to build a bilingual speech recognition system.
Fast video face detection was recently described in \cite{tran2021facial}.

The current paper extends prior methods 
    through the combination of YOLO with
    AM-FM representations.
Firstly, YOLO is used to process RGB images for face detection.
Secondly, FM images, characterized by higher instantaneous frequencies, are used
        with LeNet5 to remove non-group faces that were falsely detected by YOLO.
Thirdly, LeNet5 is used to remove false positives from the back-of-the-head classifier. 

We summarize the rest of the paper in three additional sections.
The proposed method is summarized in section \ref{sec:method}. 
The results are given in section \ref{sec:results}.
Concluding remarks are given in section \ref{sec:conclusion}.

\begin{figure*}[!t]
	\centering
	\includegraphics[width=1\textwidth]{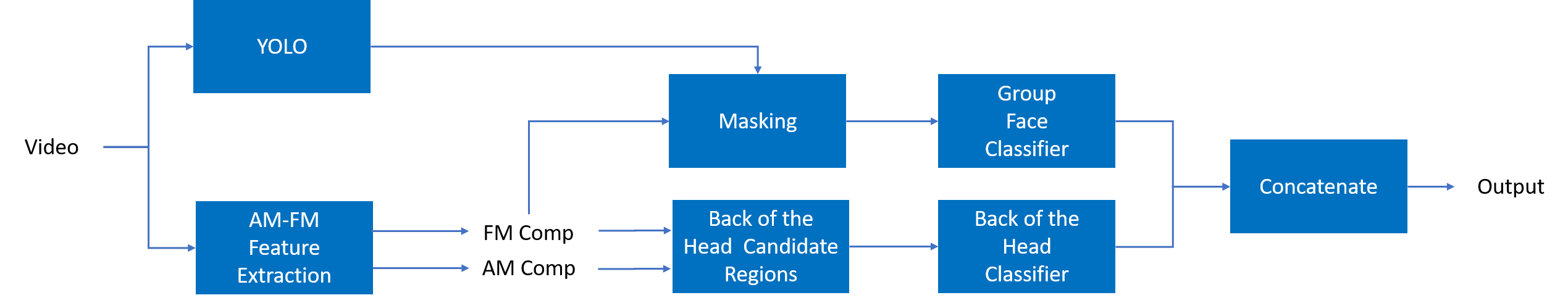}
	\caption{Student Group Detection System.} 
	\label{fig:HeadDetSystem}
\end{figure*}

\begin{figure*}[!t]
	\centering
	(a)~\includegraphics[width=0.3\textwidth]{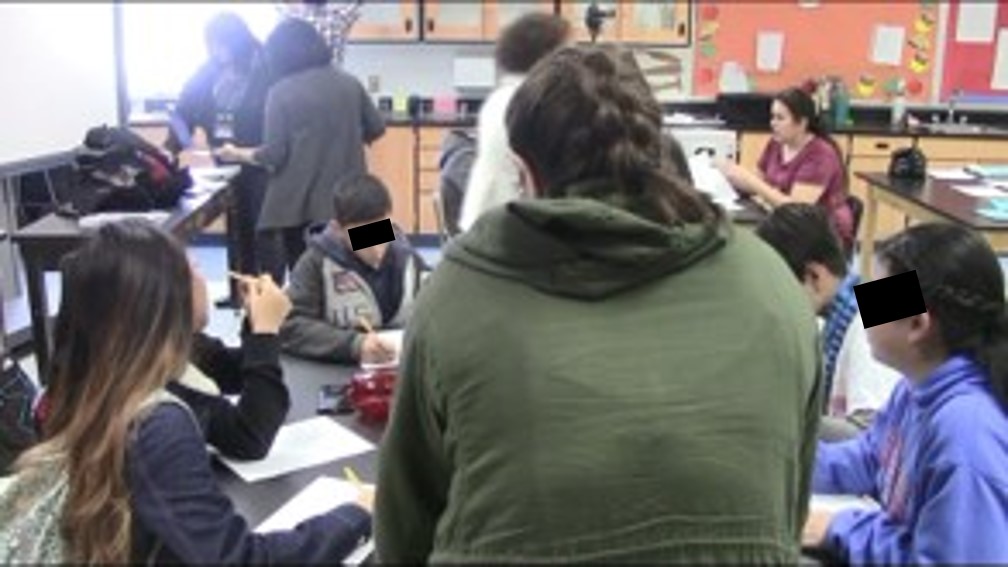}~
	(b)~\includegraphics[width=0.3\textwidth]{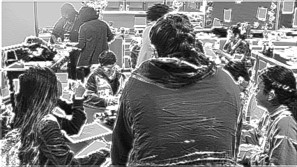}~
	(c)~\includegraphics[width=0.3\textwidth]{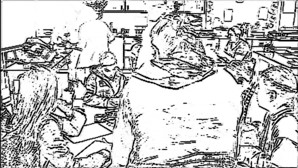}~
	\caption{AM-FM representation for classroom environment.
		(a) Classroom image.
		(b) AM component.
		(c) FM component} 
	\label{fig:AM-FM samples}
\end{figure*}

\section{Method}\label{sec:method}
We present a system diagram of the entire system in Figure \ref{fig:HeadDetSystem}. 
YOLO V3 is used for face detection.
The detected faces are further processed in
   combination with the AM-FM components as described below.

AM-FM components are extracted from the grayscale (Y-component)
     using  dominant component analysis (DCA) estimated using a 54-channel Gabor filterbank
     as described in \cite{shi2018robust}.
Using DCA, the input image frame is approximated by:
     $ I(x, y) \approx a(x, y) \cos \varphi (x, y)$
where   $ a(x, y)$ denotes the AM component and
      $ \cos \varphi (x, y)$ denotes the FM component.
Figure \ref{fig:AM-FM samples} shows the extracted AM-FM components.

The FM image is masked by the results of the YOLO face detector.
We apply this step to extract the FM components over
   students within the desired group as well as other groups.
FM components over the faces of the closest group
    will exhibit lower frequency components than the higher frequency
    components associated with distant faces from other groups.
To detect the group faces, we thus apply a simple, LeNet-based classifier 
    \cite{lecun1998gradient}
    on the extracted FM components over $100 \times 100$ pixel regions.

The AM-FM components are also used to detect the hair
    and back-of-the-head candidate regions described in \cite{shi2018robust}.
A LeNet based classifier is used to detect the back-of-the-heads against
    background detections as detailed in \cite{shi2018robust}.
For each video frame, we detect the entire group
    by concatenating the results from the face and back-of-the-head classifiers.            
      

\section{Experimental Data and Results}\label{sec:results}
The proposed methodology was trained and tested on
    digital videos recorded through actual classroom
    implementations of the Advancing Out-of-school Learning in Mathematics and Engineering (AOLME) program.
The videos depicted a variety of different learning environments with rich background
    activities and several background groups.
    
For training the YOLO face detector, we used 1000 faces and 1200 non-face images from student groups 
    extracted from 54 different videos.
Among the selected face images, we used 70\% of the 
    images for training and 30\% for validation.

For training the group face classifier, refer to Table \ref{rejectBG_table}. 
The dataset was generated from the same 54 videos from 13 different groups. 
As summarized in Table \ref{rejectBG_table}, the augmented dataset contained
    about 70,000 group face images and 70,000 non-group face images.

\begin{table}[!h]
	\caption{\label{rejectBG_table}
		Group faces classifier training, validation, and testing.
		The numbers include seven-fold data augmentation performed
		using random rescaling, cropping, rotating, and flipping.}
	\begin{center}
		\begin{tabular}{lll}
			\toprule %
			& \textbf{Group Faces}      
			& \textbf{Non-group Faces} \\ \midrule
			\textbf{Training}   & 39,232 & 39,259  \\
			\textbf{Validation} & 16,813 & 16,825  \\
			\textbf{Testing}    & 14,011 & 14,021  \\
			\textbf{Total}      & \textbf{70,056} & \textbf{70,105}  \\  \midrule
			\textbf{AUC Score}           & \multicolumn{2}{c}{0.97}          \\
			\textbf{Accuracy}            & \multicolumn{2}{c}{97.5\%}           \\       \bottomrule
		\end{tabular}
	\end{center}
\end{table}
     
\begin{table}[!h]
	\caption{\label{rejectNoBack_table}
		Back-of-the-head classifier training and testing.
		The numbers include seven-fold data augmentation performed
		using random rescaling, cropping, rotating, and flipping.
	}
	\begin{center}
		\begin{tabular}{lll}
			\toprule %
			& \textbf{Back-of-the-Heads}   & \textbf{Other} \\ \midrule
			\textbf{Training+Validation} & 22,768          & 22,800 \\
			\textbf{Testing}             & 5,710           & 5,682  \\
			\textbf{Total}      & \textbf{28,478}  & \textbf{28,482}   \\ \midrule
			\textbf{AUC Score}           & \multicolumn{2}{c}{0.97}          \\
			\textbf{Accuracy}            & \multicolumn{2}{c}{97.3\%}           \\       \bottomrule
		\end{tabular}
	\end{center}
\end{table}

Table \ref{rejectNoBack_table} describes the training and validation dataset for 
    the back-of-the-head classifier.
The dataset uses 56,000 frames from the 54 videos.
The dataset was used to train a second LeNet5 classifier to remove
     false positives when detecting the back-of-the-head regions.

\begin{figure}[!t]
	\centering
	\includegraphics[width=0.48\textwidth]{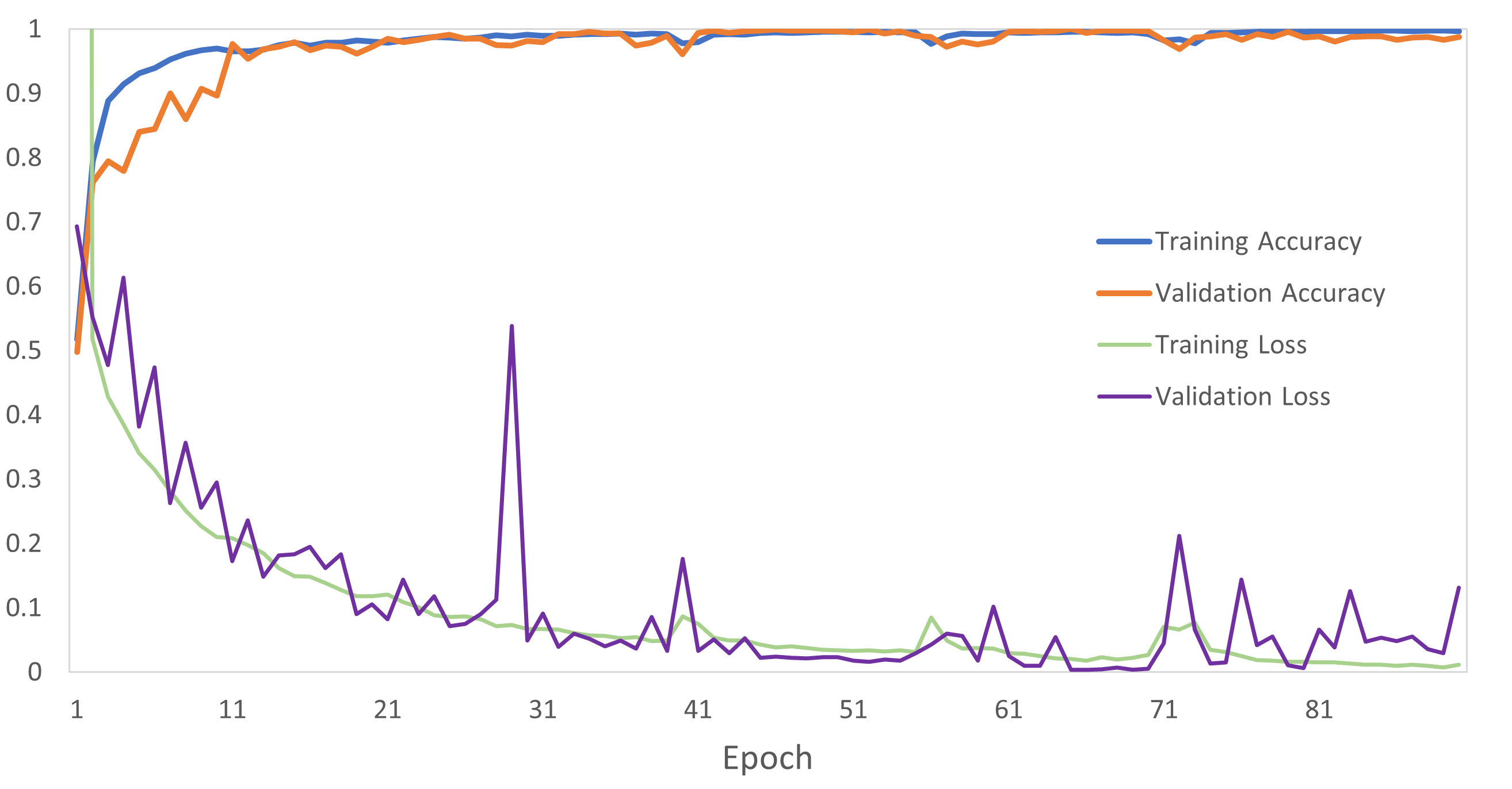}
	\caption{Group faces classifier training.} 
	\label{fig:BG_Lenet5}
\end{figure}

\begin{figure}[!t]
	\centering
	\includegraphics[width=0.48\textwidth]{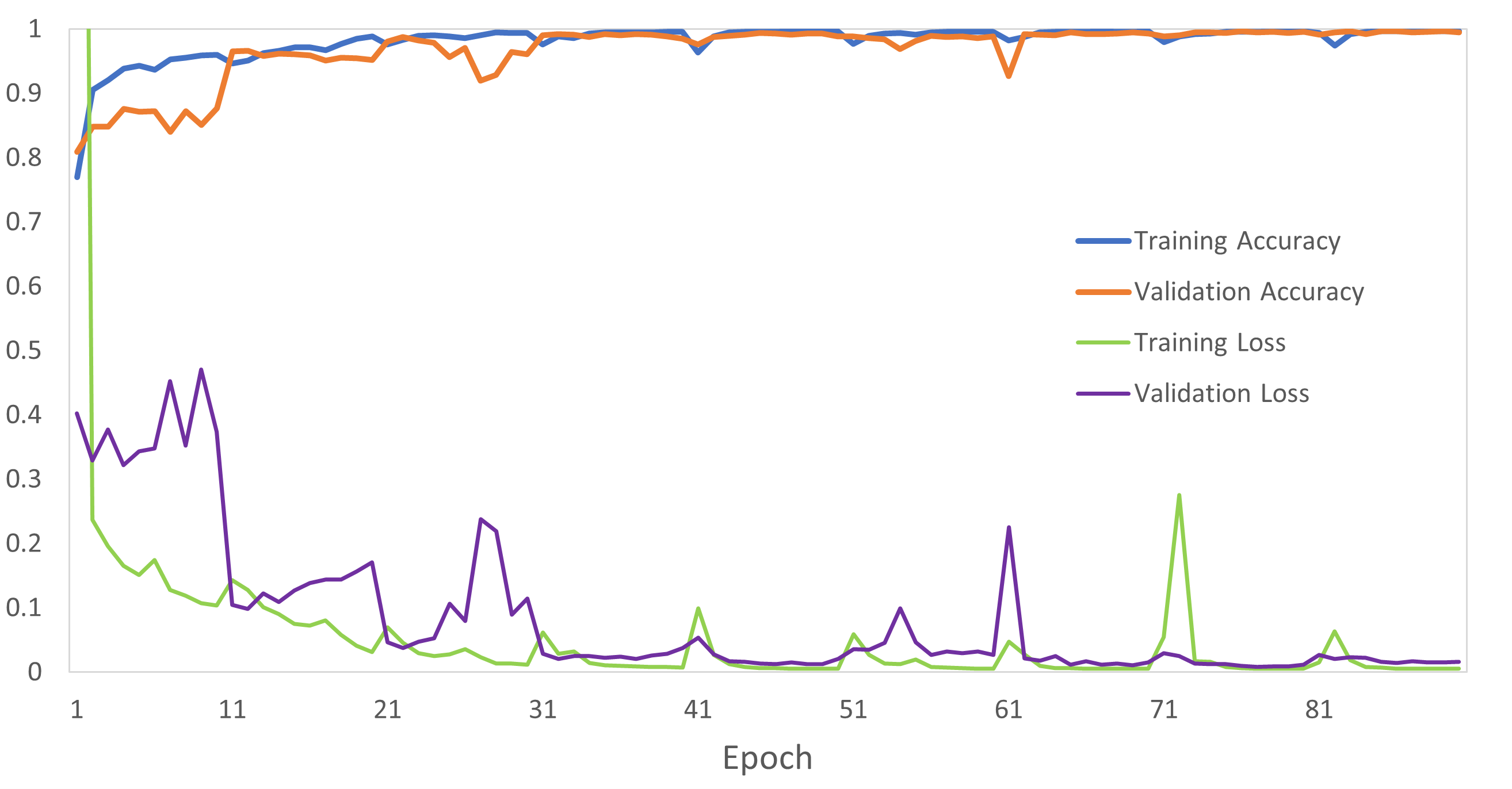}
	\caption{Back-of-the-head classifier training.} 
	\label{fig:NoBack_Lenet5}
\end{figure}

\begin{figure}[h]
	\centering
	\includegraphics[width=0.4\textwidth]{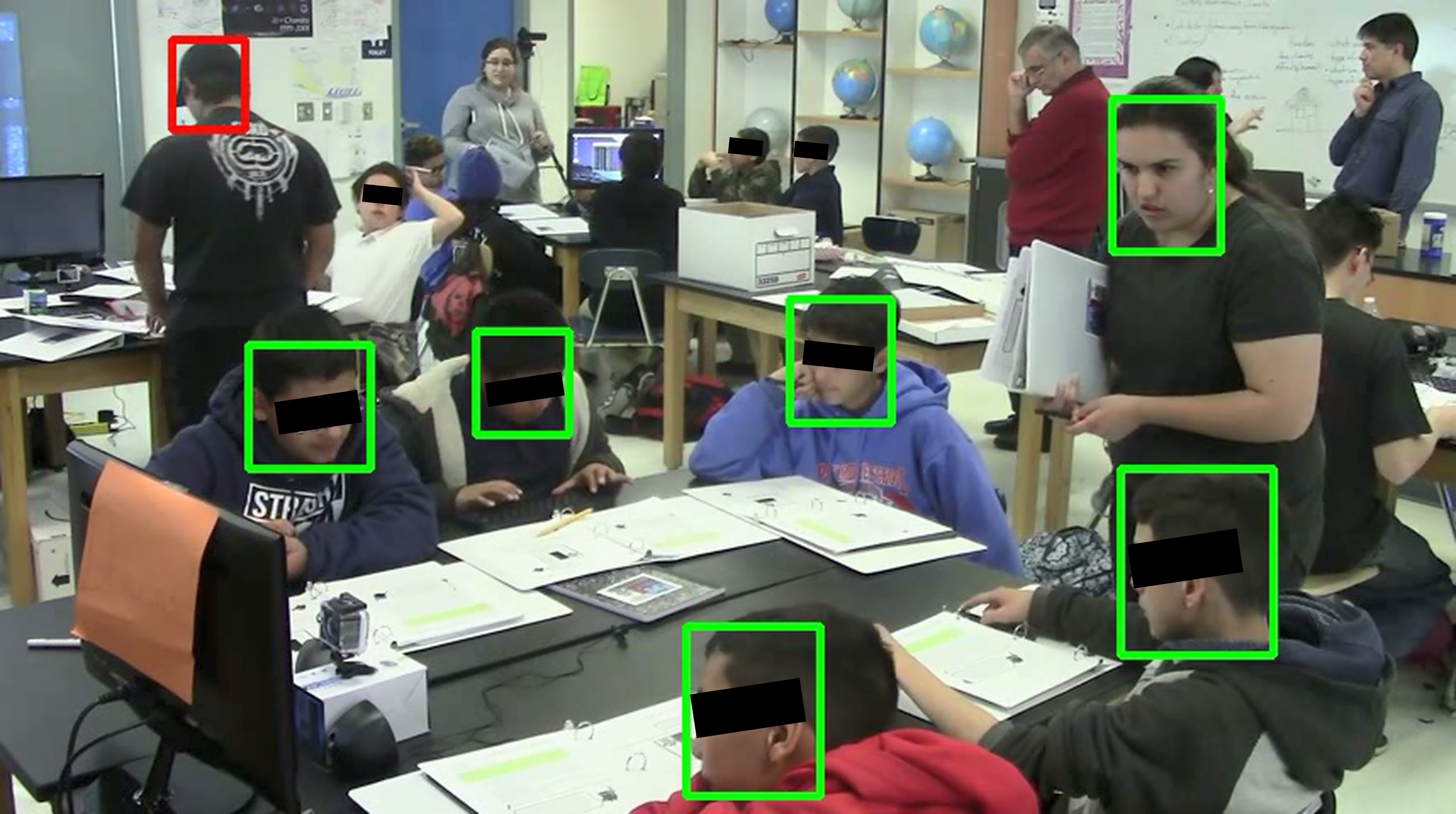}\\~\\
	\includegraphics[width=0.4\textwidth]{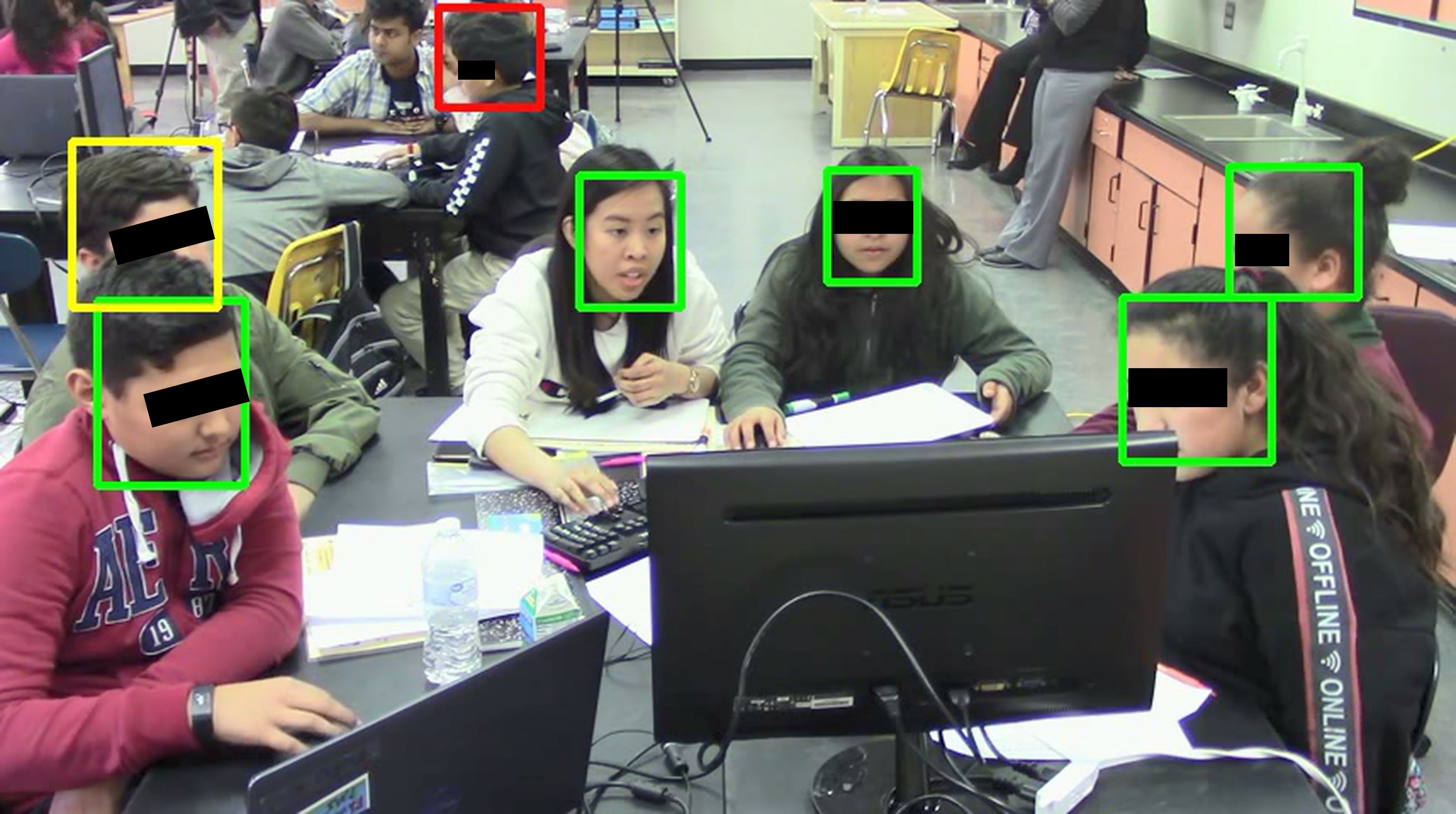}
	\caption{Head detection system results. 
		True positives are bounded by green boxes.
		False positives are bounded by red boxes.
		False negatives are bounded by yellow boxes.} 
	\label{fig:successful and failed cases}
\end{figure}

\begin{figure*}[h]
	\centering
	(a)~\includegraphics[width=0.3\textwidth]{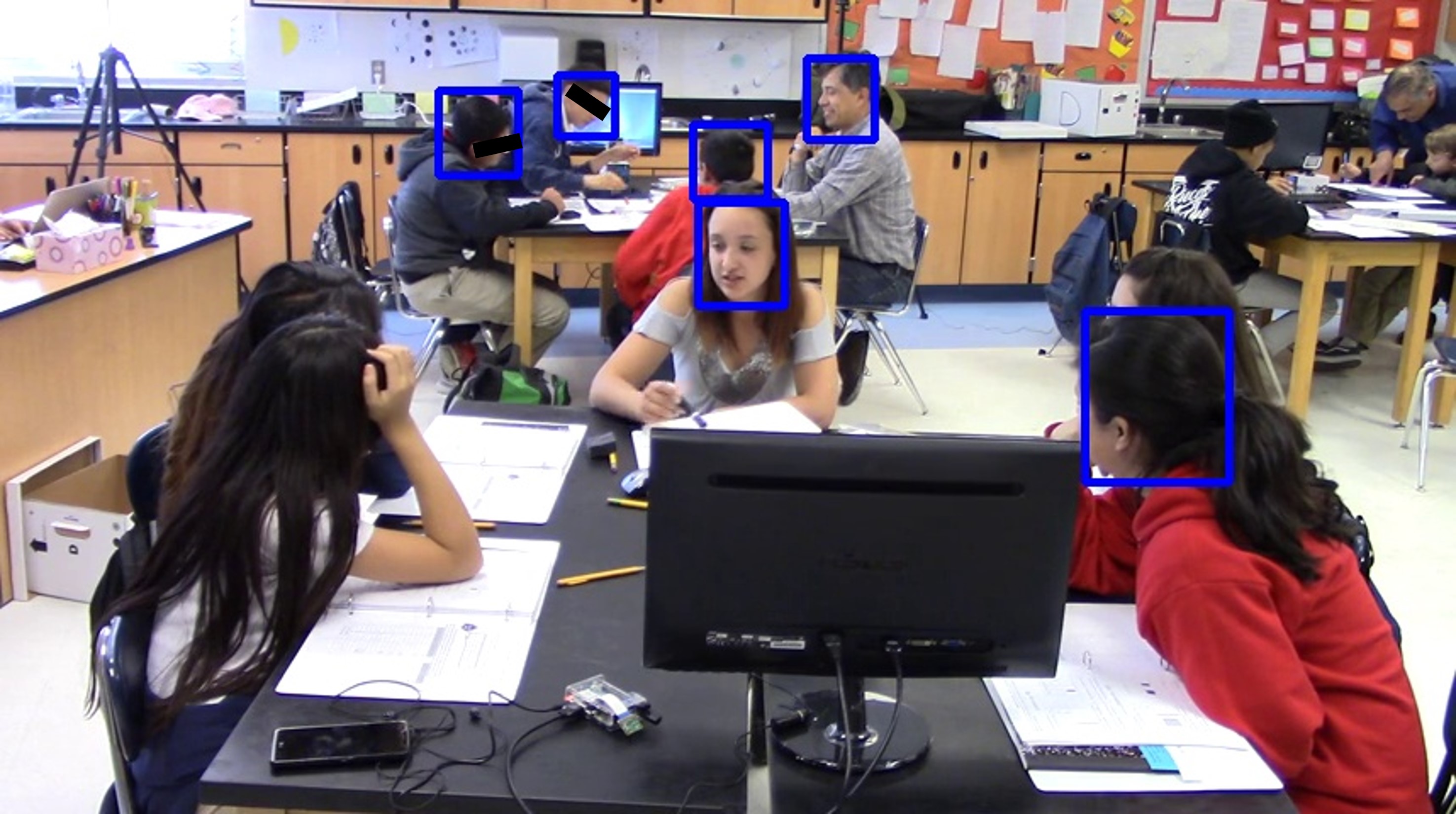}~
	(b)~\includegraphics[width=0.3\textwidth]{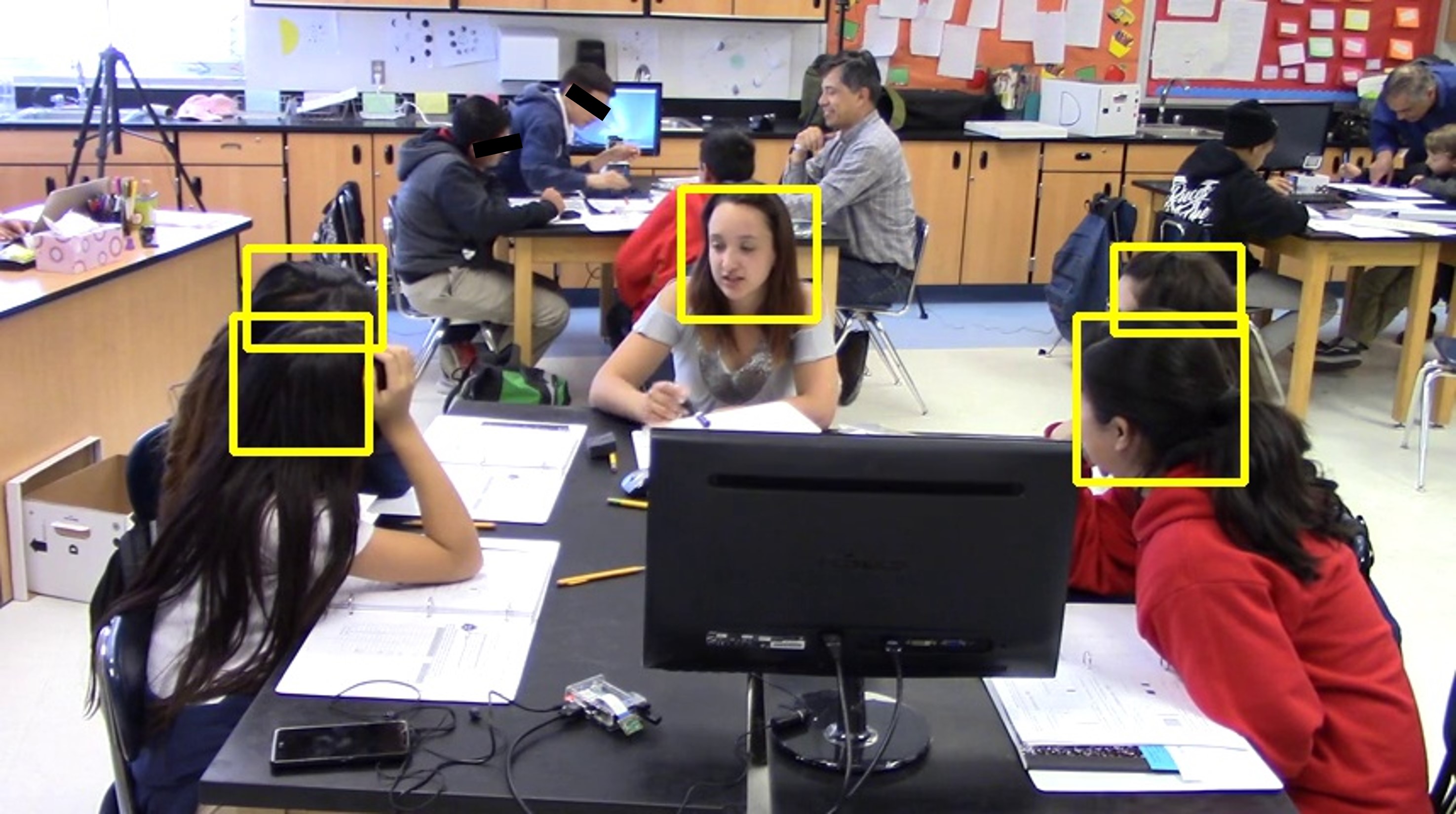}~
	(c)~\includegraphics[width=0.3\textwidth]{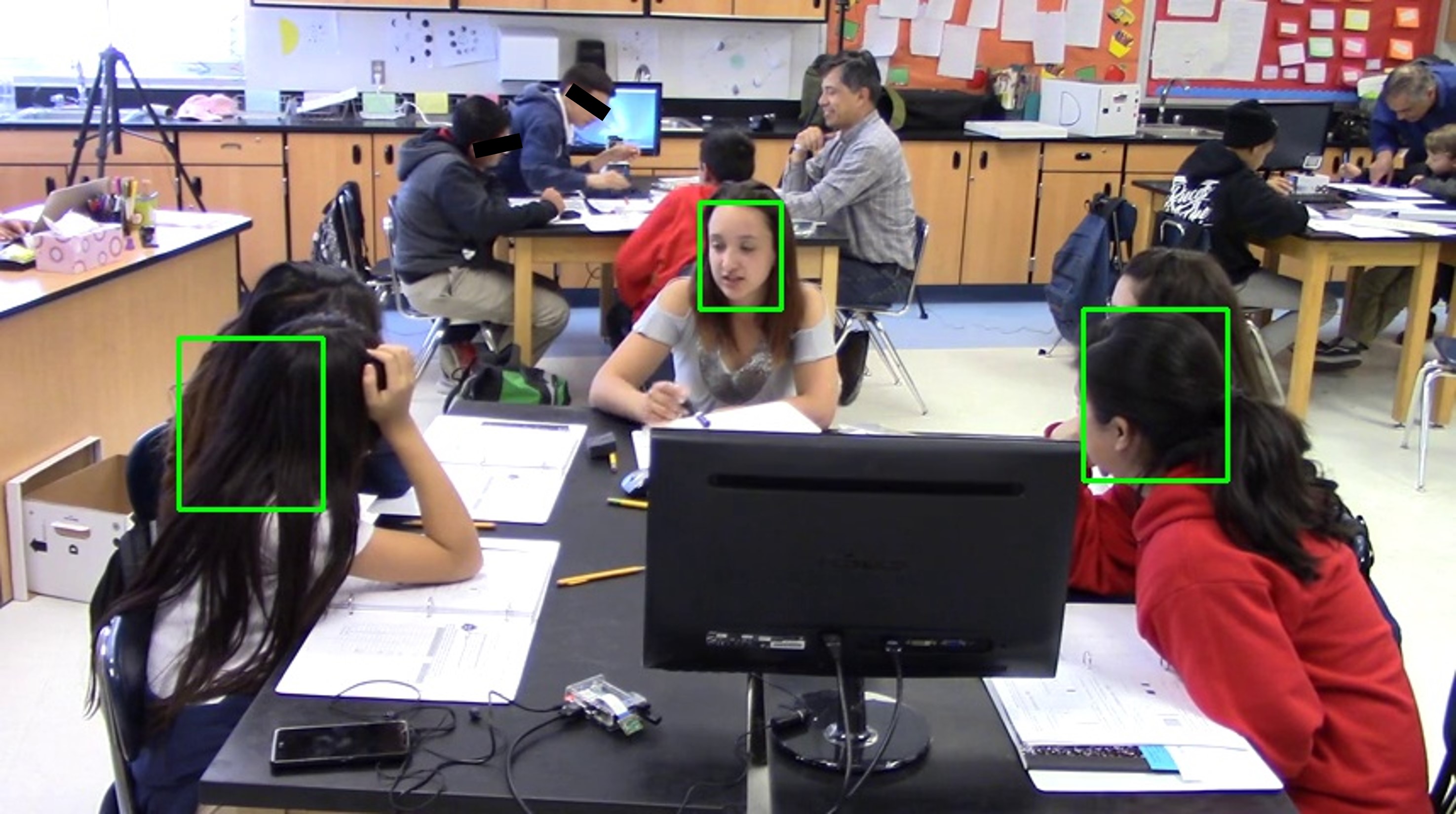}\\[0.05 true in]
	(d)~\includegraphics[width=0.3\textwidth]{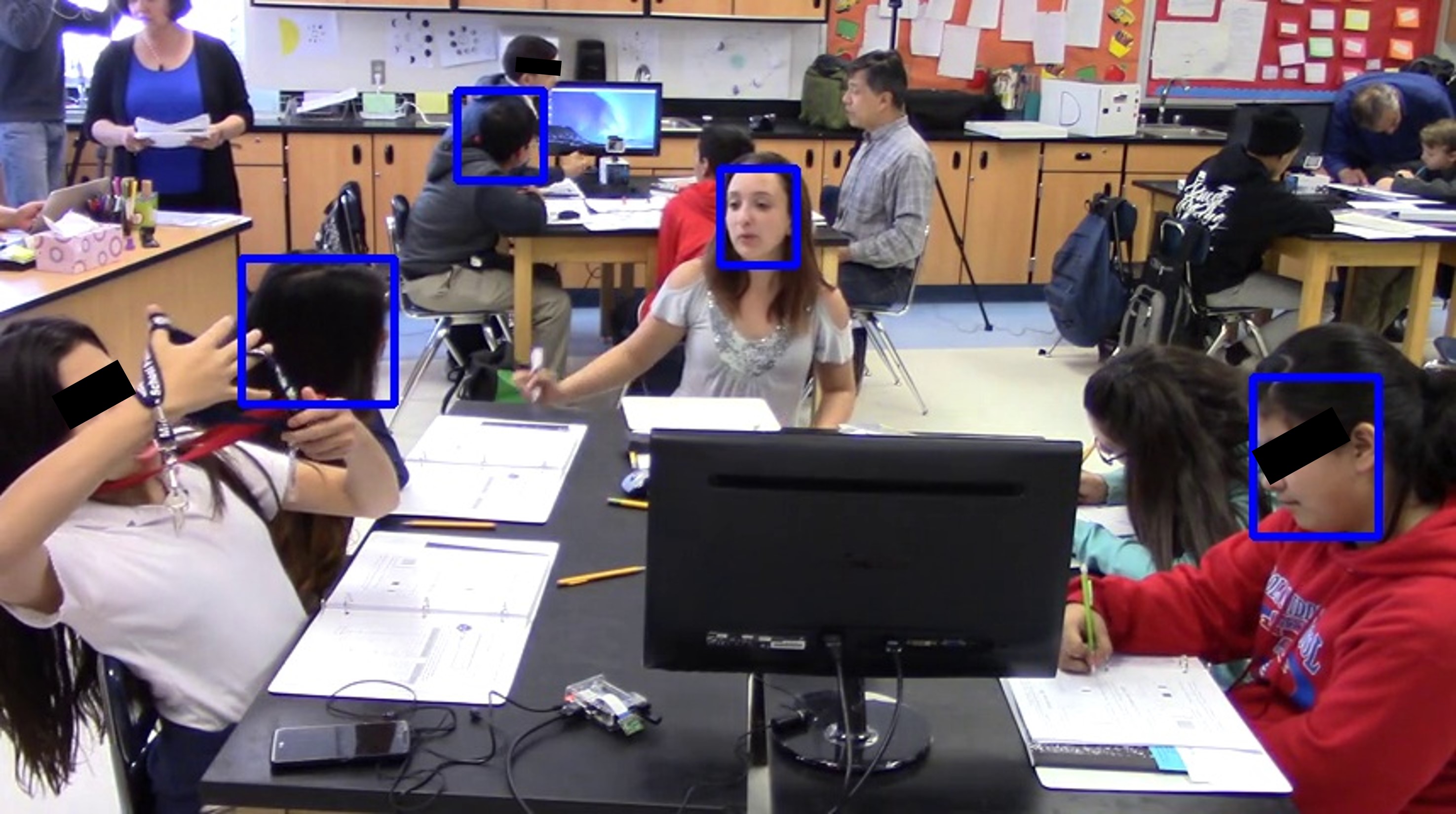}~
	(e)~\includegraphics[width=0.3\textwidth]{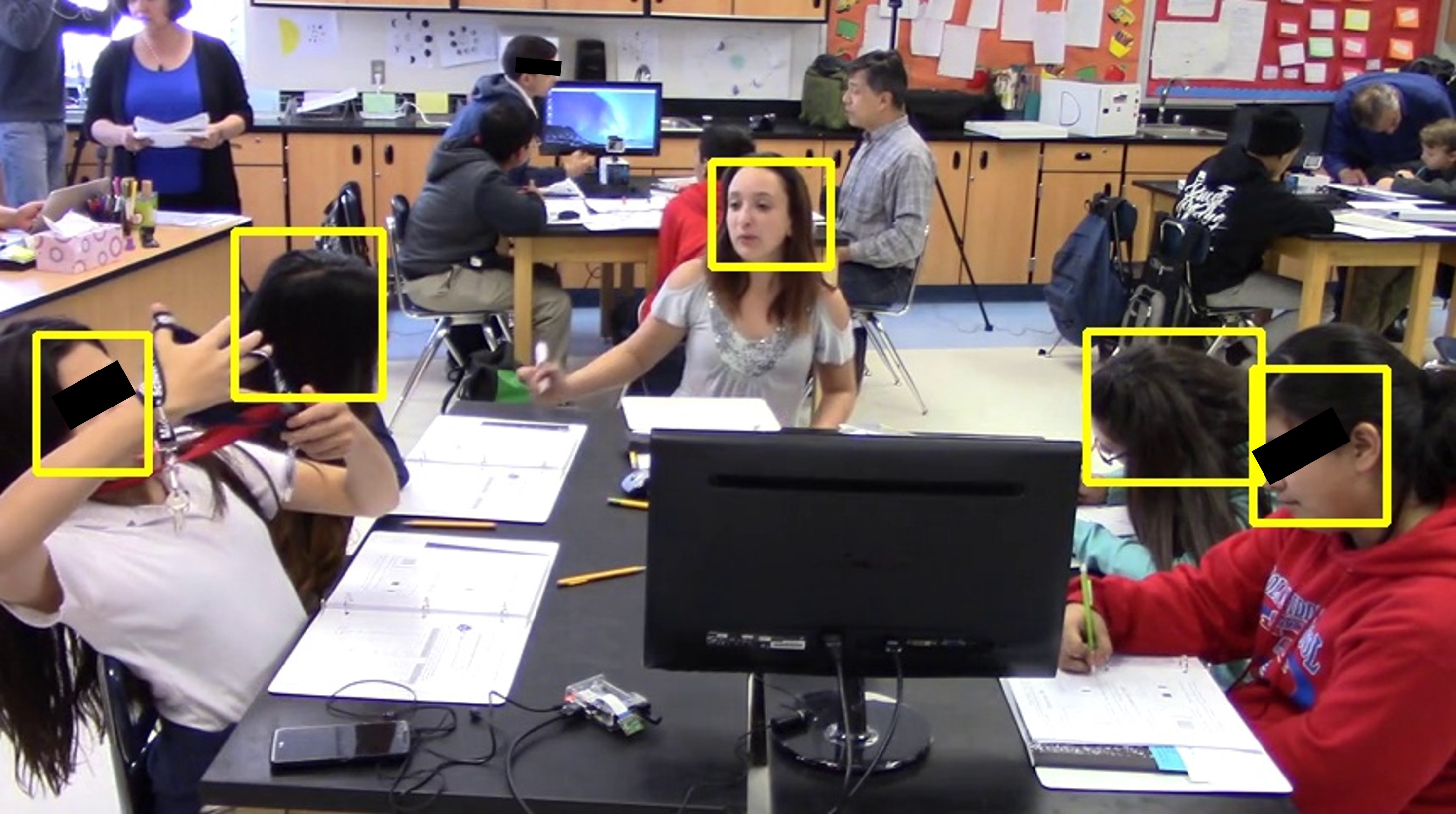}~
	(f)~\includegraphics[width=0.3\textwidth]{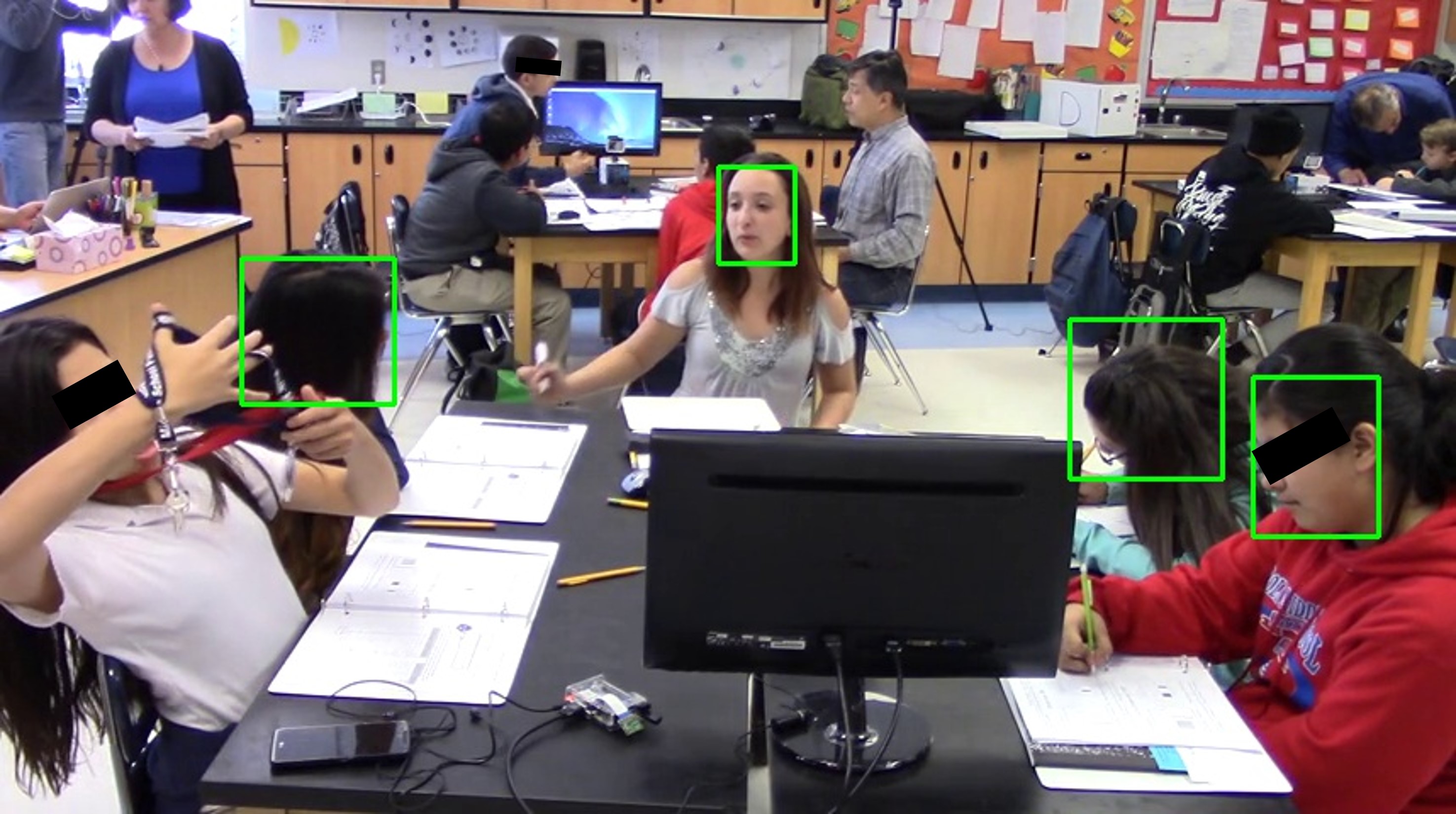}\\[0.05 true in]
	\caption{Examples from Group Detection Results. 
		Left column   ((a) and (d)) shows the results of YOLO V3.
		Middle-column ((b) and (e)) shows the ground truth.
		Right-column  ((c) and (f)) shows the results of the proposed method.} 
	\label{fig:comparison}
\end{figure*}

For the two LeNet5 classifiers, we allocated 70\% for training and 30\% for validation. 
An independent dataset, 20\% of the total images, was used for testing the final model in each case.
The training datasets include a seven-fold data augmentation performed 
    using random rescaling, cropping, rotating, and flipping.
The training and validation accuracies are provided in  Figures \ref{fig:BG_Lenet5} and
   \ref{fig:NoBack_Lenet5}. According to the table, we get over 97\% AUC score and accuracy for each model.

We used a new set of four long videos from different student groups for the final testing.
The video results are given in Table \ref{results_table}.
For successful detection, we require the intersection over union (IOU) score to be at least 0.6.
From the results, it is clear that the proposed approach outperformed the use of YOLO V3 alone.

We show examples of true positives, false positives, and false negatives in Fig. \ref{fig:successful and failed cases}.
False positives are associated with out-of-group detection.
False negatives are due to occlusion.

Figure \ref{fig:comparison} displays the comparative examples, YOLO V3 only, ground truth, and our proposed method.
From the results, it is clear that YOLO V3 cannot differentiate between the in-group and out-of-group faces.
The use of the AM-FM components allows us to remove the out-of-group faces, as shown
in the right column of Figure \ref{fig:comparison}.
  
\begin{table*}[h]
	\caption{\label{results_table}
		Comparative results for student group detection over four videos.
		TP, FP, FN refer to true positives, false positives, and false negatives,
		respectively.
		F1 scores are given for each video and each method.    
		The videos represent different student groups.
	}
	\begin{center}
		\begin{tabular}{lllllllll}
			\toprule %
			\textbf{Video} & \textbf{Length in minutes} 
			& \textbf{Labeled Students} & \textbf{Method}          
			& \textbf{Detected Students} & \textbf{TP} & \textbf{FP} & \textbf{FN} & \textbf{F1}   \\ \toprule
			\multirow{2}{*}{\textbf{V1}} & \multirow{2}{*}{96 minutes}  & \multirow{2}{*}{1,627,320}   & \textbf{YOLO}            & 1,915,935                  & 1,153,959   & 761,976     & 124,527     & 0.72          \\
			&             &             & \textbf{Proposed Method} & 1,397,790                  & 1,183,630   & 214,160     & 344,640     & \textbf{0.81} \\ \midrule
			\multirow{2}{*}{\textbf{V2}} & \multirow{2}{*}{85 minutes} & \multirow{2}{*}{887,700}     & \textbf{YOLO}            & 1,274,429                  & 723,283     & 551,146     & 12,153      & 0.72          \\
			&             &             & \textbf{Proposed Method} & 847,250                    & 728,110     & 119,140     & 110,140     & \textbf{0.86} \\ \midrule
			\multirow{2}{*}{\textbf{V3}} & \multirow{2}{*}{117 minutes} & \multirow{2}{*}{1,063,300}   & \textbf{YOLO}            & 792,291                    & 720,762     & 715,29      & 321,159     & 0.79          \\
			&              &                 & \textbf{Proposed Method} & 819,700                    & 745,880     & 73,820      & 293,640     & \textbf{0.80} \\ \midrule
			\multirow{2}{*}{\textbf{V4}} & \multirow{2}{*}{108 minutes} & \multirow{2}{*}{1,139,850}   & \textbf{YOLO}            & 1,212,963                  & 839,450     & 373,513     & 120,252     & 0.77          \\
			&             &                  & \textbf{Proposed Method} & 950,210                    & 859,290     & 90,920      & 242,850     & \textbf{0.84} \\ \bottomrule
		\end{tabular}
	\end{center}
\end{table*}

\section{Conclusion}\label{sec:conclusion}
The paper presents a method for detecting groups of students using multiple image representations.
The effective combination of YOLO V3 with AM-FM representations provides for improved results.
Our current research is focused on face recognition and talking activity detection.

\section*{Acknowledgment}
This material is based upon work supported by the National Science Foundation under Grant No. 1613637, No.1842220 and, No.1949230.

\bibliographystyle{IEEEtran}
\bibliography{bare_conf}

\end{document}